\documentclass[twocolumn,amsmath, amssymb, aps, pra,reprint, longbibliography,nofloatfix,footinbib,superscriptaddress]{revtex4}

\usepackage{multirow}
\usepackage[utf8]{inputenc}
\usepackage[urlcolor=blue, colorlinks=true,citecolor=blue]{hyperref}

\usepackage{natbib}
\usepackage{comment}
\usepackage{algpseudocode}
\usepackage{algorithm}

\usepackage[letterpaper,top=2cm,bottom=2cm,left=3cm,right=3cm,marginparwidth=1.75cm]{geometry}

\usepackage{tikz}
\usetikzlibrary{quantikz}
\usepackage{amsfonts}
\usepackage{graphicx}
\usepackage{parskip}

\usepackage{graphicx}
\usepackage{amsmath,bm}

\usepackage[normalem]{ulem}
\usepackage{bbold}
\usepackage{dsfont}
\usepackage{xcolor}
\usepackage{float}
\usepackage{booktabs} 
\usepackage{array}
\usepackage{multirow}
\usepackage{ragged2e} 
\usepackage{tabularx}
\usepackage{adjustbox}

\usepackage{soul}

\begin{document}
\title{A supervised hybrid quantum machine learning solution\\ to the emergency escape routing problem}

\author{Nathan Haboury}
\affiliation{Terra Quantum AG, Kornhausstrasse 25, 9000 St.~Gallen, Switzerland}

\author{Mo Kordzanganeh}
\affiliation{Terra Quantum AG, Kornhausstrasse 25, 9000 St.~Gallen, Switzerland}

\author{Sebastian Schmitt}
\affiliation{Honda Research Institute Europe GmbH, Carl-Legien-Straße 30, 63073 Offenbach am Main, Germany}

\author{Ayush Joshi}
\affiliation{Terra Quantum AG, Kornhausstrasse 25, 9000 St.~Gallen, Switzerland}

\author{Igor Tokarev}
\affiliation{Terra Quantum AG, Kornhausstrasse 25, 9000 St.~Gallen, Switzerland}

\author{Lukas Abdallah}
\affiliation{Terra Quantum AG, Kornhausstrasse 25, 9000 St.~Gallen, Switzerland}

\author{Andrii Kurkin}
\affiliation{Terra Quantum AG, Kornhausstrasse 25, 9000 St.~Gallen, Switzerland}

\author{Basil Kyriacou}
\affiliation{Terra Quantum AG, Kornhausstrasse 25, 9000 St.~Gallen, Switzerland}

\author{Alexey Melnikov}
\affiliation{Terra Quantum AG, Kornhausstrasse 25, 9000 St.~Gallen, Switzerland}

\begin{abstract}
Managing the response to natural disasters effectively can considerably mitigate their devastating impact. This work explores the potential of using supervised hybrid quantum machine learning to optimize emergency evacuation plans for cars during natural disasters. The study focuses on earthquake emergencies and models the problem as a dynamic computational graph where an earthquake damages an area of a city. The residents seek to evacuate the city by reaching the exit points where traffic congestion occurs. The situation is modeled as a shortest-path problem on an uncertain and dynamically evolving map.  We propose a novel hybrid supervised learning approach and test it on hypothetical situations on a concrete city graph.  This approach uses a novel quantum feature-wise linear modulation (FiLM) neural network parallel to a classical FiLM network to imitate Dijkstra's node-wise shortest path algorithm on a deterministic dynamic graph. Adding the quantum neural network in parallel increases the overall model's expressivity by splitting the dataset's harmonic and non-harmonic features between the quantum and classical components. 
The hybrid supervised learning agent is trained on a dataset of Dijkstra's shortest paths and can successfully learn the navigation task. 
The hybrid quantum network improves over the purely  classical supervised learning approach by 7\% in accuracy. 
We show that the quantum part has a significant contribution of 45.(3)\% to the prediction and that the network could be executed on an ion-based quantum computer. The results demonstrate the potential of supervised hybrid quantum machine learning in improving emergency evacuation planning during natural disasters.
\end{abstract}

\maketitle 
\section{introduction}

Natural disasters like earthquakes can result in devastating effects, including loss of life and property damage \cite{jaiswal2009estimating, badal2002prognostic}. Emergency evacuation procedures are critical in such scenarios, and optimizing these procedures is essential for saving lives \cite{10.48550/arxiv.2202.12505}. 
One of the most common modes of transportation during emergency evacuations is cars, and it is important to ensure that the routes taken by these vehicles are safe and efficient.
The standard road network, however, can be heavily affected by earthquakes through dynamic effects like land deformation, collapsing buildings or debris \cite{ering2020effect, song2022two, nabian2018deep, xu2019analysis, nishino2012evaluation}.
Such effects can be modelled and applied in traffic simulation using sophisticated probabilistic models \cite{costa2020application, wu2022post}. 
Using such models, a complete solution for medical rescue, including route planning, which considers collapsed buildings, was proposed in \cite{xu2022joint}.
This study, however, excludes the consideration of traffic capability or capacity due to the challenges involved in obtaining post-earthquake travel data.
The central challenge of optimization-based methods~\cite{10.1007/978-3-319-60042-0_15, 10.1016/j.proeng.2014.04.023,10.3844/jcssp.2015.330.336, 10.1109/icemms.2010.5563425, osanlou2022planning} is the complexity of large-scale problems, in particular on evolving (dynamic) networks~\cite{10.1111/j.1475-3995.2008.00649.x}.

Dijkstra's algorithm effectively finds the optimal path on a static graph, and while algorithms like A*~\cite{astar} might offer faster alternatives, Dijkstra's is the only one with an optimality guarantee~\cite{dijkstras}. However, this algorithm struggles to find the shortest path in an evolving and uncertain situation. Therefore, it is necessary to adapt the algorithm for graphs with dynamically changing edge weights by rerunning it every time the graph is modified. We refer to this as the node-wise Dijkstra's algorithm.
Furthermore, Dijkstra's algorithm (node-wise or otherwise) requires global knowledge of the graph. This would require accurate, up-to-date graph information, which is not always feasible to obtain in reality, as pointed out in \cite{xu2022joint}. In particular, in this problem, this information would require perfect evolving traffic information at each time.
This impracticality incentivizes a solution that uses only local information and is robust to unreliable traffic data in the graph. We introduce a hybrid quantum machine learning approach that only requires local information and aims to mimic the node-wise Dijkstra's algorithm in terms of path quality \cite{dijkstras, 10.2307/3689390} on a dynamic graph.

Applying quantum technologies to machine learning has shown much potential in recent years~\cite{dunjko2018machine,qml_review, Biamonte_2017, Cerezo_2022, cerezo_challenges_2022-1,whitepaper}. Hybrid quantum machine learning techniques, which combine quantum computing and classical machine learning, have emerged as promising approaches to tackle industrial problems~\cite{ic50,thales,tq-vw-QML,senokosov2023quantum,evonik,kurkin2023forecasting}. This paper explores the potential of hybrid quantum machine learning for optimizing emergency escape plans for cars during natural disasters. The study aims to additionally provide a general blueprint for using hybrid quantum machine learning for an industrial-scale problem by including analyses to address the circuit efficiency and the quantum processing unit (QPU) integration.

We use supervised learning (SL) and train on decisions of the node-wise Dijkstra's algorithm in a simulated earthquake scenario. The focus is on the dynamic environment with evolving natural disaster and where the traffic builds up  at the city's exit locations. 

For this case, we demonstrate the potential of hybrid quantum machine learning in improving the efficiency of emergency evacuation plans during natural disasters.  Sec.~\ref{sec:problem} outlines the problem statement, first in general, and then models it by creating an evolving environment as a dynamic computational graph in Sec.~\ref{sec:problem_modelling}.  In Sec.~\ref{sec:methods}, we describe the SL approach, the specifics of the hybrid machine learning model used, and the results obtained. Sec.~\ref{sec:analysis} provides a comprehensive practical and theoretical analysis of the quantum model, its contribution to the inference, prospects of QPU integration, and model efficiency.  The latter was subdivided into three types of analysis -- ZX calculus, Fourier embedding analysis, and Fisher expressivity -- all of which were considered when developing the final hybrid model. Finally, Sec.~\ref{sec:conclusion} summarizes the results. 

\section{The Emergency Escape Routing Problem} \label{sec:problem}
The research objective of the problem is to develop an effective strategy for rapid evacuation of a city during emergencies such as earthquakes, fires, or floods.  Cars are considered a form of transportation to exit the city. Hence traffic congestion has to be bypassed. This study focuses on earthquake emergencies.

\subsection{Problem Setting}
\label{sec:problem_setting}

\begin{figure*}[t]            
\centering    
     \includegraphics[width=1.0\textwidth]{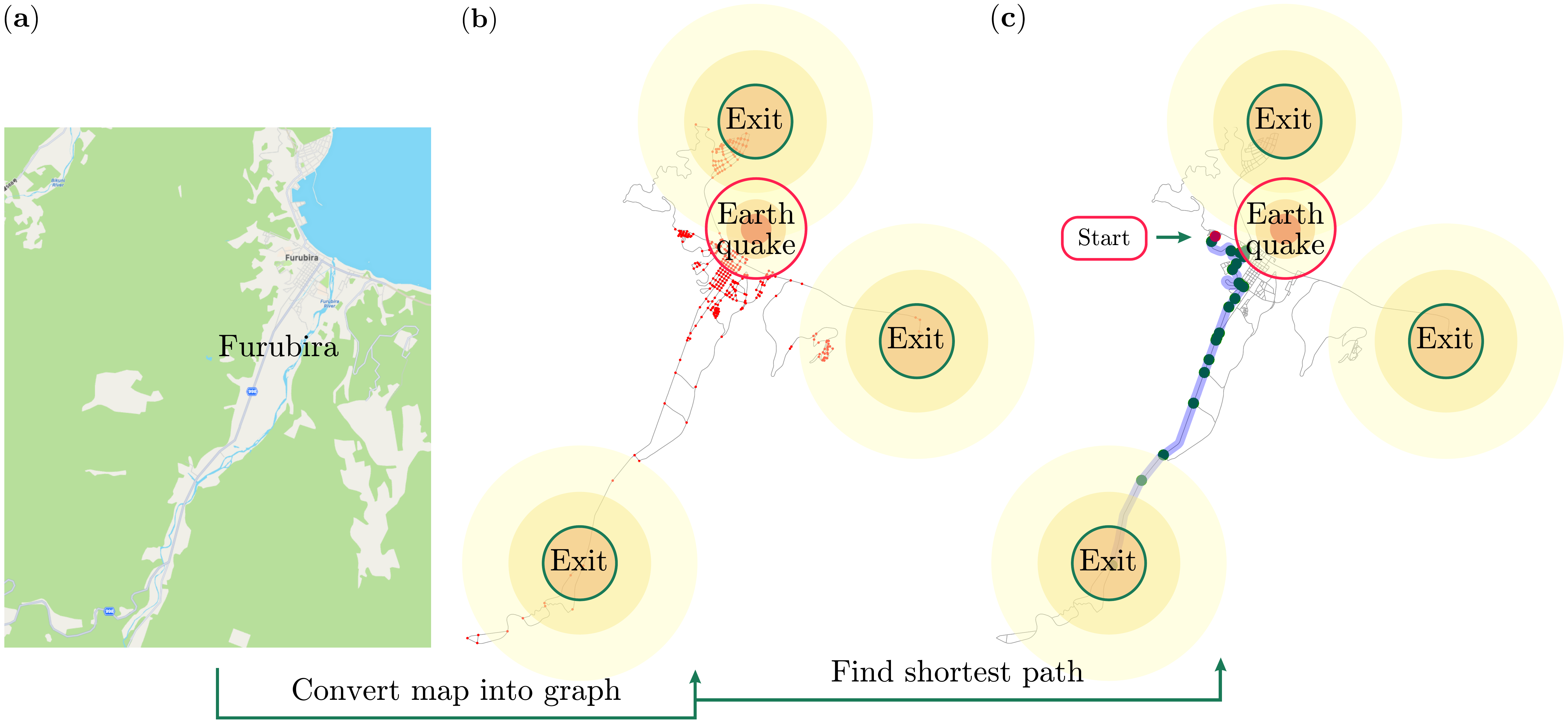}
    \caption{The mathematical abstraction of the emergency escape routing problem during an earthquake. The map of Furubira was investigated by converting it to a computational graph. Each node represents a crossroad where a decision on the direction is required. The three exit points on the map are expected to produce traffic congestion.  To model this, we created three expanding concentric circles that increasingly impacted the edges around them. We also modeled a dynamic earthquake that had affected an area centered around an epicenter.  The description of the problem comprises navigating this dynamic graph to the closest exit point in a time-efficient manner.}
    \label{map_raw}
\end{figure*}

The emergency escape routing problem is considered for a specific map depicted in Fig.~\ref{map_raw}(a). On the map, there are several predefined exit locations where all traffic should go to. Additionally, there is one earthquake area. The traffic flow nearby the exit points increases over time, resulting in higher travel times in this area. 
In addition, it is assumed that an earthquake constantly affects roads after it occurs, increasing travel time in its vicinity. Each car has access to the resulting up-to-date traffic information for its immediate surroundings. Depending on the current traffic, a car can change its route planning at every time step. A car is considered evacuated when it reaches an exit point, leaving the city region. The objective is to obtain a route for evacuating cars, which minimizes travel time. Given the current location of a car, its respective optimal evacuation route to an exit point leaving the city region should be found.

\subsection{Map Data Source and Preparation}
\label{sec:real_world_data}

The ultimate goal is to solve the routing problem on examples of real-world cities. This is because some design logic is used in city layouts, intended to be captured in our model. Using the Python OSMnx package~\cite{osmnx}, any selected region of a map can be converted into a graph that represents a realistic city or regional scenario. Fig.~\ref{map_raw}(a) shows an example of such a selected map region, where the resulting graph is shown in Fig.~\ref{map_raw}(b), with nodes as dots representing intersections and edges as lines representing street segments. The resulting undirected graph has 357 nodes and 549 edges. The minimum and maximum node degree is two and five, respectively. Each edge of the graph represents a road segment and has a weight reflecting the travel time along that segment.

The graph data includes the speed limit for most streets, depending on the road type. For streets where this is missing, a nominal value can be added. The travel time is calculated using the speed limit and road length attributes. The travel time down a given road is randomly sampled from this nominal value based on a Gaussian distribution. This can be used to simulate changing traffic conditions. Exit points are sampled uniformly at some strategic places on the graph, e.g. exterior of the city/village and close to major highways. The graph will evolve around the exit points, where we simulate traffic evolution. In addition to the exit point, an earthquake also affects the graph. In Sec.~\ref{sec:problem_modelling}, we explain how these changes affect the graph. 

A dataset of many problem instances (graphs with different conditions) is generated. For each instance, we have a different epicenter with random coordinates. The starting point also is chosen randomly for each instance. Three exit points are defined for the chosen map of the Furubira region. For each instance, one of these is randomly chosen as exit points. Fig.~\ref{map_raw}(c) is an example of a path found using Dijkstra's algorithm. The algorithm will return the shortest path for a given starting node for a weighted graph.

\subsection{Mathematical Problem Abstraction}
\label{sec:problem_modelling}

This section proposes a mathematical model for the effects of an earthquake and traffic flows nearby exit nodes defined in Section \ref{sec:problem_setting}. The earthquake initially has a static impact by increasing the edge weights, i.e.\ travel times, at the beginning of the simulation. After that, it has an ongoing dynamic effect, increasing nearby edge weights, while the area of effect also increases over time.
The road segments near the exit nodes increase the edge weights dynamically while the car moves from node to node.
To define these mechanisms, the time $t$ is introduced. It starts with $t=0$ and is increased by 1 every time a node is traveled. It, therefore, equals the current total number of steps taken.
To simulate the weight-increasing effects on the graph, three mechanisms are used. The first mechanism simulates the initial static effect of the earthquake, while the second covers its dynamically evolving effect. The third mechanism simulates the dynamics around the exit nodes, i.e.\ the ongoing traffic flow. For each step that is taken in the environment, these three mechanisms update the weights subsequently as described in Algorithm \ref{alg:cap}. 

\begin{algorithm}

\begin{algorithmic}[1]
\caption{Subsequent weight update}\label{alg:cap}

\State Initialize graph
\State t = 0
\State Update graph weights according to the initial earthquake effect mechanism 
\Repeat
\State Update graph weights according to ongoing earthquake effect mechanism 
\State Update graph weights according to ongoing traffic effect mechanism 
\State Travel to the next node chosen by the model
\State t += 1
\Until {Exit node is reached}
\end{algorithmic}
\end{algorithm}

\begin{figure*}[t]            
\centering    
    \centering
     \includegraphics[width=1\textwidth]{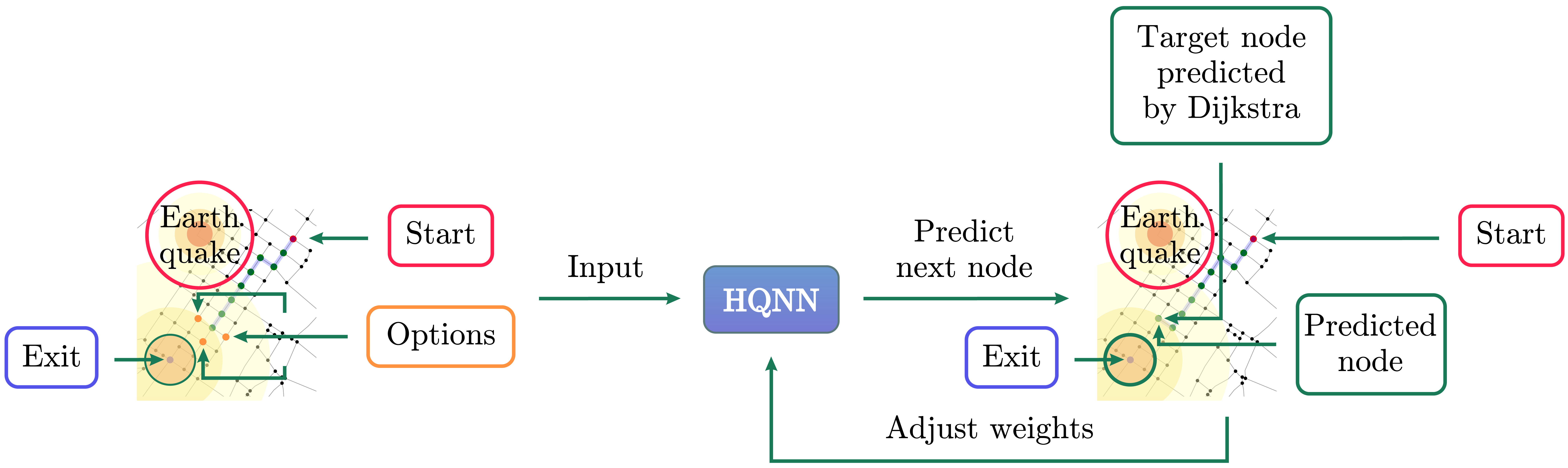}
    \caption{The supervised learning approach to the emergency escape routing problem. The hybrid quantum neural network (HQNN) trains to imitate the node-wise Dijkstra's algorithm.}
    \label{fig:model_SL_RL}
\end{figure*}

The first two mechanisms simulate the initial and ongoing effects of the earthquake, respectively, depending on the earthquake epicenter.
An earthquake epicenter is defined as having a circular area of influence that increases over time $t$. The earthquake's radius is denoted as the damage radius,  which grows over time as $r_{\text{epi}} = 0.5 + \sqrt{0.0002 \times t}$. The algorithm first simulates the initial earthquake effect and increases the edge weights once at the time $t=0$. Then it covers the ongoing effect of the earthquake and increases the edge weights in the area of effect over time. These increases in the edge weights depend on the respective Euclidean distance between the center of the edge and the center of the effect (exit nodes or the earthquake epicenter).  For the earthquake, this distance is denoted as $d_{\text{epi}}$. The shorter this distance, the stronger the increase.
The update function for each edge weight $w$ for the initial increase is defined as:
\[
    w \gets 
\begin{cases}
    w \times 5 ,& \text{if } d_{\text{epi}} \leq 0.3~r_{\text{epi}} \\
    w \times 2 ,& \text{if }  0.3~r_{\text{epi}}   < d_{\text{epi}} \leq 0.75~r_{\text{epi}} \\
    w \times 1.3 ,& \text{if }  0.75~r_{\text{epi}} < d_{\text{epi}} \leq r_{\text{epi}} \\
    w,              & \text{otherwise.}
\end{cases}
\]
And the subsequent time evolution is given by:
\[
    w \gets 
\begin{cases}
    \min \{w \times \sqrt{0.003 \times t + 1}~,~5\} ,\\ \qquad\text{if } d_{\text{epi}} \leq 0.3~r_{\text{epi}} \\
    \min \{w \times \sqrt{0.002 \times t + 1}~,~4\} ,\\ \qquad\text{if }  0.3~r_{\text{epi}} < d_{\text{epi}} \leq 0.75~r_{\text{epi}} \\
    \min \{w \times \sqrt{0.001 \times t + 1}~,~3\} ,\\ \qquad \text{if }  0.75~r_{\text{epi}}  < d_{\text{epi}} \leq r_{\text{epi}} \\
    w, \qquad \text{otherwise,}
\end{cases}
\]
where at each step, all $w$'s are updated and used as the baseline values for the next step. To simulate the traffic flow, the third mechanism dynamically increases the edge weights near exit nodes while traveling from the start to the exit node. Similar to the earthquake simulation, an exit point has a circular area of effect with radius $r_{\text{exit}}$. 
This radius increases over time, and within the area of effect, all edge weights are increased as well. Each edge's increase depends on its Euclidean distance to the exit node coordinates, denoted as $d_{\text{exit}}$. The lower this distance, the larger the increase. The corresponding update function for each edge weight $w$ is defined as:
\[
    w \gets 
\begin{cases}
    \min \{w \times \sqrt{0.03 \times t + 1}~,~5\} ,\\ \qquad\text{if } d_{\text{exit}} \leq 0.5~r_{\text{exit}} \\
    \min \{w \times \sqrt{0.02 \times t + 1}~,~4\} ,\\ \qquad\text{if }  0.5~r_{\text{exit}} < d_{\text{exit}} \leq 0.75~r_{\text{exit}} \\
    \min \{w \times \sqrt{0.01 \times t + 1}~,~3\} ,\\ \qquad \text{if }  0.75~r_{\text{exit}}  < d_{\text{exit}} \leq r_{\text{exit}} \\
    w, \qquad \text{otherwise}
\end{cases}
\]
with $r_{\text{exit}} = \sqrt{0.00075 \times t}$.

\section{Setup and Methods} \label{sec:methods}

To solve the emergency escape routing problem, an SL-based method, shown in Fig.~\ref{fig:model_SL_RL}, is proposed. It consists of a hybrid quantum neural network (HQNN) that iteratively chooses the next node based on the car's current state and map. The SL model is trained on a dataset with labels generated by node-wise Dijkstra's algorithm. This way, the HQNN approximates Dijkstra's algorithm while only accessing a limited portion of the map.

\subsection{Data Engineering}

The input data fed to the SL approach in Fig.~\ref{fig:model_SL_RL} consists of the earthquake coordinates, the start and destination node coordinates, the adjacent edges of the current node with their respective edge weights, which encode the travel time along each edge. We also add the betweenness centralities of each edge.

The edge betweenness centrality is a measure in network analysis that quantifies the number of times a particular edge acts as a bridge along the shortest path between two other nodes. In the context of the given scenario, it is calculated based on the original city graph without considering the impacts of earthquakes or traffic conditions.

Additionally, two heuristic indicators are introduced to represent global information. 
These are added to the input data for each adjacent edge and represent the types of questions that a human driver might ask when deciding where to navigate from that node, which are: \\
1) am I getting close to the destination? \\
2) am I heading toward the destination?

The answers to the questions are encoded in the Euclidean distance and cosine distance, respectively, between the current node and the target node. The Euclidean distance is defined for two nodes $p=(p_x,p_y)$ and $q=(q_x,q_y)$ as:
\begin{equation*}
    d(p,q) =  \sqrt{ (q_{x}-p_{x})^2+(q_{y}-p_{y})^2} 
\end{equation*}

The cosine distance is:
\begin{equation*}
    \cos(\theta) = \frac{\mathbf{A} \cdot \mathbf{B}}{\left\Vert\mathbf{A}\right\Vert \left\Vert\mathbf{B}\right\Vert}
\end{equation*}
For a node $p$ and a neighbor node $q$ we define $A$ and $B$ as: $A = (q_{x}-p_{x}, q_{y}-p_{y})$ $B = (\text{exit}_{x}-p_{x},\text{exit}_{y}-p_{y})$.

The input variables for the HQNN model are detailed in Table~\ref{tab:inputs}. The features $4$ to $8$ are specific to each edge and thus are included for each edge adjacent to the current node in the model input. Given the maximum node degree in the graph is 5, the total input contains 36 values:
\begin{align*}
    \text{Input} = [x_{\texttt{epi}},y_{\texttt{epi}},x_{\texttt{start}},y_{\texttt{start}},x_{\texttt{dest}},y_{\texttt{dest}}, \\
    x_{\texttt{edge}_1},y_{\texttt{edge}_1}, w_1, e_1, d_1, c_1, \\
    ...,\\
    x_{\texttt{edge}_5},y_{\texttt{edge}_5}, w_5, e_5, d_5, c_5 ]
\end{align*}

These comprise features $1$ to $3$, as well as five series of features $4$ to $8$ for each adjacent edge from Table~\ref{tab:inputs}.
If a node has less than five adjacent edges, zero padding is employed on the input vector to extend it to a length of 36 in order to keep the model input dimension constant.

\begin{table}[ht]
\centering
\caption{Model input summary}\label{tab:inputs}
\setlength{\tabcolsep}{8pt}
\begin{tabular}{ll}
1) Earthquake coordinates & $x_{\texttt{epi}},y_{\texttt{epi}}$\\
2) Start node coordinates& $x_{\texttt{start}},y_{\texttt{start}}$ \\
3) Destination coordinates &$x_{\texttt{dest}},y_{\texttt{dest}}$ \\     
4) End of edge coordinates&   $x_{\texttt{edge}_n},y_{\texttt{edge}_n}$    \\
5) Required travel time &     $w_n$    \\
6) Edge betweenness centrality& $e_n$ \\
7) Euclidian distance&     $d_n$  \\
8) Cosine distance & $c_n$ \\
\end{tabular}
\end{table}

\subsection{Model Evaluation Metrics}

We evaluate the models based on two metrics. 
The first metric characterizes the effectiveness, which quantifies whether a model can find an escape route, i.e.\ the probability that the model succeeds in finding a path from the start to the exit node in the graph.  
This is done by sampling random start and exit node pairs and evaluating the arrival rate, i.e.\ the probability of finding a connecting path between these two nodes as   
\begin{equation*}
   \textrm{ Arrival rate} = \frac{\textrm{No.\ instances path found}}{\textrm{No.\ sampled node pairs}} 
\end{equation*}

The second metric evaluates the path quality and is called accuracy, which counts the total travel time along a path relative to the node-wise Dijkstra result.
The total travel time is calculated as the sum of all edge weights for a given path, and thus the accuracy is given by 
\begin{equation*}
   \textrm{Accuracy}= 1 - \left|1 - \frac{\sum_{\textrm {weight }} \textrm { path }_{\textrm {Dij }}}{\sum_{\textrm {weight }} \textrm { path }_{\textrm {model }}}  \right|\:,
\end{equation*}
where $\textrm { path }_{\textrm {Dij }}$ and $\textrm { path }_{\textrm {model }}$ are the set of edges  in the paths of the Dijkstra algorithms and the learned model, respectively. 

\subsection{Hybrid Supervised Learning Architecture}

\begin{figure*}[ht]            
\centering    
    \centering
     \includegraphics[width=1\textwidth]{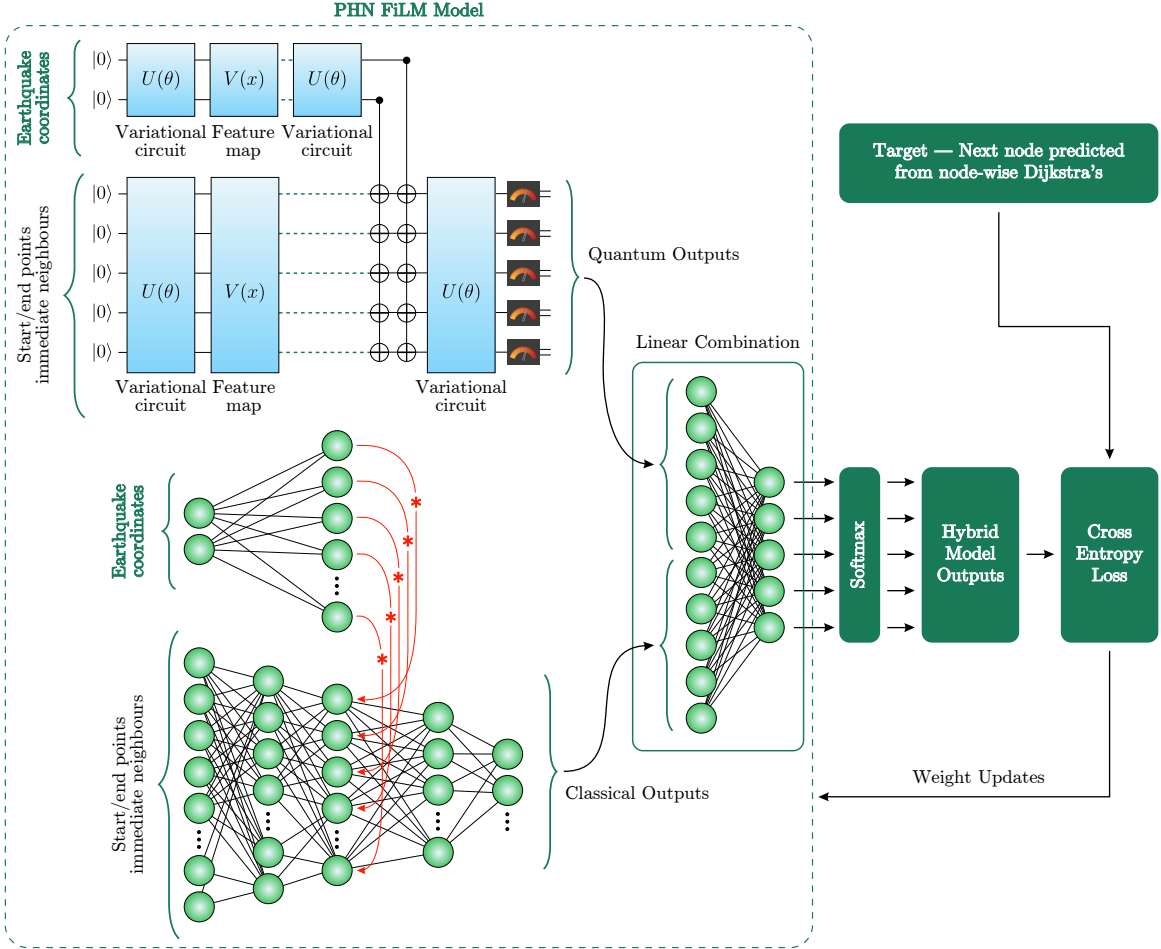}
    \caption{A diagrammatic view of the model architectures used in this work. The hybrid supervised learning approach combines the classical FiLM neural network with a similarly-built quantum neural network. The former employs FiLM layers for processing the earthquake coordinates by passing these coordinates into two fully-connected layers. The layers then become multiplicative and additive values for the main body of the neural network that processes the rest of the features. The quantum network is created similarly, where there are seven qubits in total, five of which encode the system's main features, including the features of the start, current, and end nodes of the path and information about the neighbors. The other two qubits process the earthquake coordinates using a data reuploading circuit. The latter qubits are then entangled with the main five, then a trainable layer is added, and finally, the main qubits are measured in the Z basis. These outputs are combined linearly with the outputs of the classical FiLM neural network to produce the outcome.}
    \label{fig:approaches_SL}
\end{figure*}

The choice of machine learning architecture in this work is driven by the properties of the dataset.  The dataset is produced by simulating an earthquake at randomized coordinates in the city and then collecting routing data for each earthquake simulation. Therefore, the earthquake coordinates are the same for each bunch of routing data.  Naively, this can be tackled in two ways: 1) using conditional neural networks, where for each earthquake, we train a different neural network, and 2) by adding the earthquake coordinates as new features to a slightly larger neural network. The former is resource-intensive and unable to generalize to other earthquakes, whereas the latter suffers when there is a lack of variability in the data, for example, when the number of different earthquakes is much smaller than the total number of routes.  

To counteract this, we employ feature-wise linear modulation (FiLM) neural networks~\cite{FiLM} to create a smooth and trainable conditional network. This specialized architecture is bifurcated into two primary components: the FiLM layer, which takes the earthquake coordinates as input (FiLM features), and a traditional neural network segment for the remaining features.

The FiLM layer plays a vital role in this architecture, interfacing directly with the penultimate layer of the standard neural network. It acts as a modulation agent, conducting element-wise scaling and shifting operations on the intermediate representation resulting from the parallel network. In our context, the FiLM layer exploits the earthquake coordinates and modulates the traditional neural network layer to guide the prediction of the subsequent routing node.

This modulation mechanism offered by FiLM assists in curbing the inaccuracy and resource-intensiveness of the solution if one had chosen either of the two naive paths. The FiLM layer's unique influence on subsequent layers allows for better incorporation and reflection of contextual information, such as the earthquake coordinates, thus enabling a more adaptive and accurate routing prediction.

\begin{table}[htbp]
    \centering
    \caption{Model Parameters}
    \label{tab:model_params}
    \begin{tabular}{ll>{\raggedright\arraybackslash}p{3cm}>{\raggedright\arraybackslash}p{2cm}}
        \toprule
        \hline
        \multirow{8}{*}{General} & Batch Size & \multicolumn{2}{c}{\texttt{2000}} \\
        & Input Size (main) & \multicolumn{2}{c}{\texttt{34}} \\
        & Input Size (FiLM) & \multicolumn{2}{c}{\texttt{2}} \\
        & Output Size & \multicolumn{2}{c}{\texttt{5}} \\
        & Scaling Learning Rate & \multicolumn{2}{c}{\texttt{1e-3}} \\
        & Weight Decay & \multicolumn{2}{c}{\texttt{1e-5}} \\
        & Epochs & \multicolumn{2}{c}{\texttt{100}} \\
        & Optimizer & \multicolumn{2}{c}{\texttt{Adam}} \\
        & Loss function & \multicolumn{2}{c}{\texttt{CE}} \\
        \midrule
        \hline
        \multirow{5}{*}{Classical} & Start Learning Rate Scaling Factor & \multicolumn{2}{c}{\texttt{1}} \\
        & End Learning Rate Scaling Factor & \multicolumn{2}{c}{\texttt{0.1}} \\
        & Classical Learning Rate & \multicolumn{2}{c}{\texttt{1e-3}} \\
        & Hidden Dimension & \multicolumn{2}{c}{\texttt{100}} \\
        & Activation function & \multicolumn{2}{c}{\texttt{ReLU}} \\
        & FC layers & \multicolumn{2}{c}{\texttt{3}} \\
        \midrule
        \hline
        \multirow{3}{*}{Quantum} & Number of Qubits & \multicolumn{2}{c}{\texttt{7}} \\
        & Quantum Learning Rate & \multicolumn{2}{c}{\texttt{1e-3}} \\
        & Variational Layers & \multicolumn{2}{c}{\texttt{4}} \\
        & Number of Repeats & \multicolumn{2}{c}{\texttt{1}} \\
        \bottomrule
        \hline
    \end{tabular}
\end{table}

Fig.~\ref{fig:approaches_SL} shows the architecture of the HQNN realized as a parallel hybrid network (PHN)~\cite{PHN}, which combines a classical NN and a variational quantum circuit (VQC). PHN was found to be a valuable and performant HQNN layer in industrial applications~\cite{thales,evonik,kurkin2023forecasting}. In the new PHN FiLM Model introduced in this work, the information flows in parallel in both FiLM sub-models. The hyperparameters of the PHN FiLM Model architecture are given in Table.~\ref{tab:model_params}.

The VQC is a parameterized quantum circuit that takes in the state features and the earthquake as input and outputs a list of expected values of the variational quantum states. The FiLM inputs are the earthquake coordinates, and the other features are passed to the main body of the model - see Table~\ref{tab:inputs}. This list corresponds to the likelihood of choosing each of the neighboring nodes. The circuit's architecture consists of two main parts - the FiLM and the main sections. The FiLM section consists of two qubits and accepts the earthquake coordinates using data re-uploading~\cite{data_reuploading,schuld_fourier,mo-paper} using Pauli Z rotation gates. The encoding gates are repeated five times in the circuit and interlaced with variational unitaries built using four sub-layers of Pauli X rotations and CNOT gates, known as the basic entangler layer (BEL)~\cite{pennylane-basicentanglerlayer}. This layer, denoted as $U(\theta)$ can be written as: 

\begin{align*}
    U_{\text{BEL}}(\mathbf{\theta}_l) = 
    \prod_{t=1}^{n_\text{sub-layers}} \prod_{q=1}^{n_{\text{qubits}}} 
    CX_{q,q+1}
    e^{
       -\frac{i}{2}
       \theta^{t,q}_l
       \sigma^{(q)}_X 
     },
\end{align*}

where $CX_{a,b}$ is the CNOT gate where $a$ is the control qubit, and $b$ is the target qubit (we employ cyclic conditions where each qubit index is taken modulo $n_{\text{qubits}}$). For $d$ reuploading layers, we get: 

\begin{align*}
    \ket{\psi}_{\text{FiLM}} = S_{\text{FiLM}}(x_{\text{epi}},y_{\text{epi}},\theta) \ket{0}^{\otimes n_\text{qubits}},
\end{align*}

where $S_{\text{FiLM}}(x_{\text{epi}},y_{\text{epi}},\theta)$ is given by: 

\begin{align*}
\left(\prod_{l=1}^{d} U_{\text{BEL}}(\mathbf{\theta}_l) e^{-\frac{i}{2}(\sigma^{(1)}_Z x_{\text{epi}} + \sigma^{(2)}_Z y_{\text{epi}})} \right) U_{\text{BEL}}(\mathbf{\theta}_0)
\end{align*}

The main section accepts the rest of the variables by embedding the state data into a five-qubit feature-parameter quantum depth-infused layer (QDIL) similar to Ref.~\cite{ic50}. The way this layer works is by first breaking down the main input vector $\mathbf{x^{(\text{main})}}$ into $n_{\text{subvec}}$ sub-vectors of the size of the number of qubits, using padding where necessary, such that $n_{\text{features}}\leq n_{\text{qubit}} \times n_{\text{subvec}}$. Initially, a variational layer consisting of basic entangler layers with four sub-layers is applied to the ground state. Then, a layer of Pauli Z-rotations encodes the first feature subvector, followed by another variational layer. This process is repeated multiple times until all subvectors are encoded in the system. 

\begin{align}
\ket{\psi}_{\text{main}} = S_{\text{QDIL}}(\mathbf{x},\mathbf{\theta})\ket{0}^{\otimes n_{\text{qubits}}},  \\
S_{\text{QDIL}} = \left(\prod_{l=1}^{n_{\text{subvec}}} U_{\text{BEL}}(\mathbf{\theta}_l) V(\mathbf{x}_l) \right) U_{\text{BEL}}(\mathbf{\theta}_0)
\end{align}
where $\mathbf{x}_l$ denotes the $l$'th feature subvector, $U_{\text{BEL}}(\theta_l)$ the $l$'th basic entangler layer used as the variational layer, and $V(\mathbf{x}_l)$ the $l$'th encoding layer.  The encoding layer is defined as 
\begin{align}
    V(\mathbf{x}_l) = \prod_{t=1}^{n_{\text{qubits}}} e^{-\frac{i}{2} x_{l+t}\sigma^{(t)}_Z }, 
\end{align}

where $x_{l+t}$ is the $t$'th feature in the $l$'th feature subvector, and $\sigma^{(t)}_Z$ is the Pauli-Z matrix applied to qubit $t$.  

Then, the two FiLM qubits are entangled with the main section using CNOT gates controlled using the two qubits and the NOT applied to all the qubits in the state. Finally, another variational basic entangler layer is applied to the state qubits, giving the final quantum state of the variational circuit as:

\begin{align*}
    \ket{\psi} = U^{\text{main}}_{\text{BEL}}(\theta) \left[\prod_{c=1}^{2} \prod_{t=3}^{7} CX_{c,t} \right] \ket{\psi}_{\text{FiLM}}\otimes \ket{\psi}_{\text{main}}
\end{align*}

The five main qubits are then measured in the computational basis many times to produce a list of bitstrings, each of length five. Then, using post-selection, the expectation value of each qubit was measured separately with respect to the Pauli-Z basis. To do this for a qubit $q$, the number of measurements where that qubit is $1$ is subtracted from the number of times it is $-1$ and then divided by the total number of measurements. This was repeated for every main qubit, and so 5 expectation values were generated.  This model is realized on the QMware hybrid quantum cloud \cite{benchmarking}.

\begin{figure*}[ht]            
\centering    
    \centering
     \includegraphics[width=0.8\textwidth]{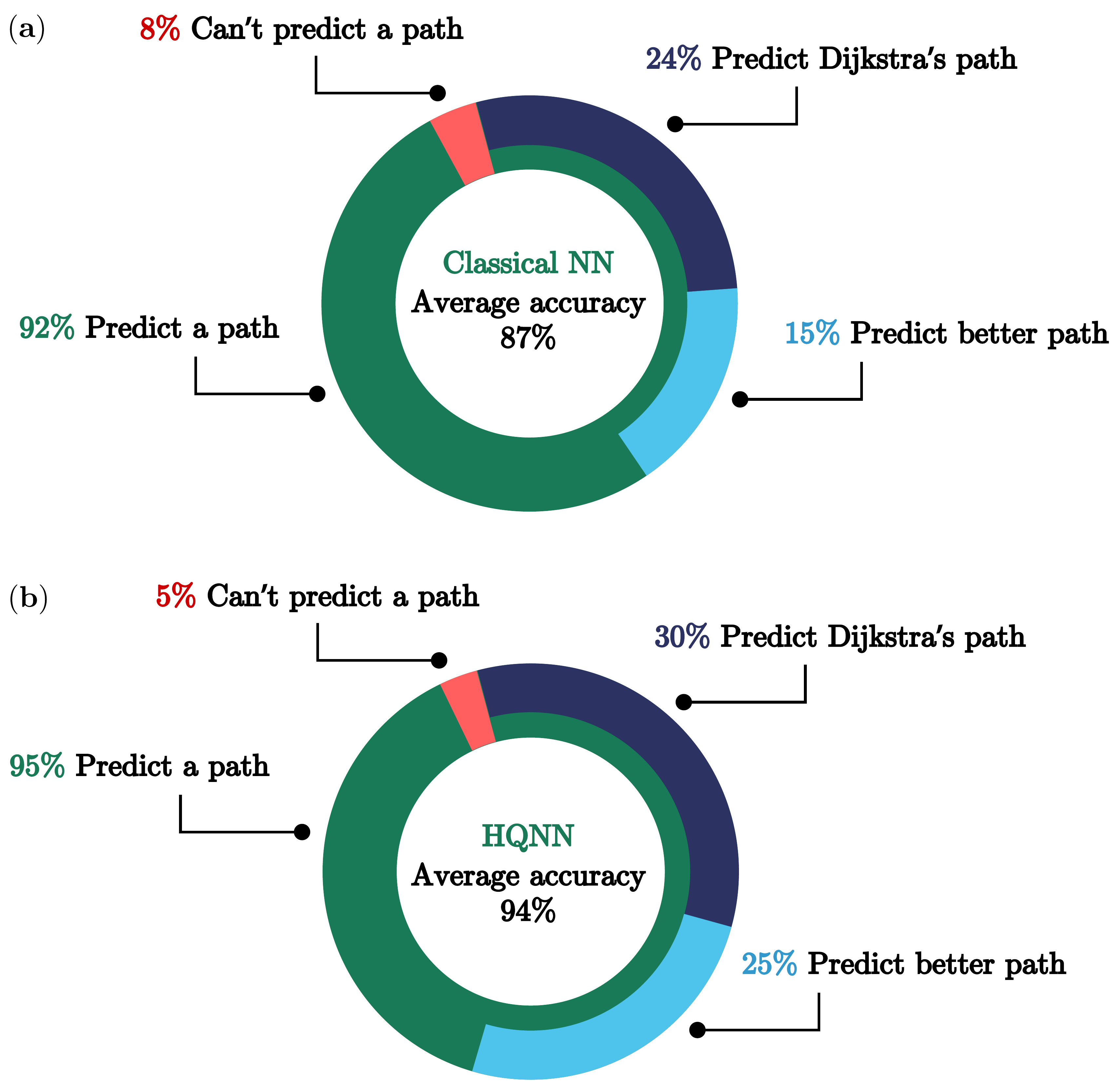}
    \caption{The results of the supervised learning approach and their comparison. (a) and (b) compare the classical NN and hybrid HQNN solutions with different metrics. The hybrid model achieved higher results in the percentage of paths predicted, but also the percentage of paths predicted that are better or the same as paths found by Dijkstra's algorithm.}
      \label{fig:results}
\end{figure*}

The classical FiLM network consists of a multi-layer perceptron (MLP) with three fully connected layers where the input layer has 36 nodes and the hidden layers have 100 nodes each.  The classical network has a ReLU activation function applied to the hidden layers, and dropout regularization with a rate of 0.5 is applied after each hidden layer to prevent overfitting. Finally, the classical network has five output nodes.  

The outputs of the quantum network are concatenated with those of the purely  classical FiLM network. 
These ten values are then passed to a fully-connected layer and reduced to five values. The outputs of this architecture are five numbers that act as the logit layer of the node classifier. Subsequently, the neighboring node corresponding to the highest number is chosen as the next node. 

\section{Results and Analysis} \label{sec:analysis}

It is desirable to explore whether a hybrid quantum approach to solving the shortest path problem could offer an improvement over classical machine learning. 

\subsection{Results}\label{sec:results}

The results of the classical NN and HQNN are presented in Fig.~\ref{fig:results}.
The classical NN model reaches an average accuracy of 87\% and an arrival rate of 95\%, showing the capability of mimicking Dijkstra's algorithm.

The HQNN model significantly improved over the classical NN model, achieving an average accuracy of 94\%. This means that, on average, the HQNN model predicts a path 7\% closer to the node-wise Dijkstra's predicted path. Furthermore, as depicted in Fig.~\ref{fig:results}, the hybrid model predicts more successful paths than the classical solution and finds more paths that are faster or equal to Dijkstra's algorithm. 

It should be noted that the trained models can predict better results than the baseline Dijkstra algorithm due to the dynamically changing environment.
Node-wise Dijkstra always chooses the next edge to travel based on the current travel times on the graph. This can lead to sub-optimal choices in case the travel times in the next time step change. The learned networks can apparently acquire knowledge on the existence of a dynamically evolving graph and make more robust choices.  

By assumption, the node-wise Dijkstra's algorithm is unavailable to the user as she only has access to the local traffic information. However, she can still run the traditional Dijkstra's algorithm on the full map, whose runtime increases with the complexity $\mathcal{O}(n_{\text{edges}}+ n_{\text{nodes}}\text{log}(n_{\text{nodes}}))$, whereas the runtime of the HQNN only depends on the number of nodes in one path which in the worst case scenario includes all nodes $\mathcal{O}(n_{\text{nodes}})$.

In addition to the practical success of the hybrid FiLM model in solving the problem, it is desirable to understand the applicability of this model in practice and the theoretical properties of the integrated parameterized quantum circuit. Sections~\ref{sec:analysis:practical}~and~\ref{sec:analysis:theory}  address these matters. 

\subsection{Practical Analysis}\label{sec:analysis:practical}
In this section, we analyze the final hybrid model in two practical ways: 1) PHN primacy and 2) QPU performance.  The former relates to the parallel nature of the network and assesses if either the VQC or the MLP has dominated the training. The latter demonstrates the technical feasibility of running this model on an ion-based quantum computer. 

\subsubsection{Primacy in the Hybrid Model} \label{sec:analysis:primacy}

\begin{figure}[t]            
\centering
\includegraphics[width=1\linewidth]{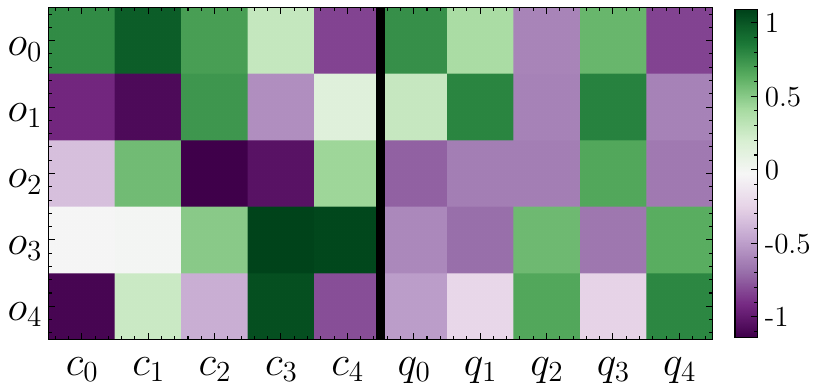}
    \caption{Matrix of weights for the final fully connected layer of the PHN, integrating the outputs of the VQC and MLP. Our analysis reveals that there is no discernible dominance of any specific component. Furthermore, both classical and quantum networks exhibit comparable contributions to the outcome. Interestingly, the quantum component exhibits smoother transitions between its submatrix elements, indicating a more uniform distribution of contributions from all quantum outputs than the classical submatrix.}
    \label{fig:primacy}
\end{figure}

PHN suffer from the problem of primacy~\cite{PHN}. This occurs when the network decides to discard the output of either the MLP or the VQC and achieve a sub-optimal minimum rather than train to include both and reach lower minima. To assess the contribution of each sub-network, we consider the weights of the final fully-connected layer of the PHN, which connect the outputs of the VQC and the MLP to the model outputs. In our problem, this corresponds to a $5\times 10$ matrix.  Fig.~\ref{fig:primacy} shows this matrix for a fully-trained hybrid model.  The matrix provides two insights: 1) the general values of the weights are similar between the VQC and the MLP, and therefore no primacy is observed, and 2) the quantum side of the matrix exhibits smoother transitions, whereas the MLP side is more irregular. 

To quantify the relative contribution of the VQC, we use the relative Frobenius norm $\alpha_q = \frac{\lVert W_{q,o}  \rVert}{\lVert W_{q,o}  \rVert + \lVert W_{c,o}  \rVert}$, where $\lVert X \rVert = \sqrt{\text{Tr}(X^2)} $ and $W_{x,y}$ refers to the weight matrix connecting the $x$ neurons (in this case classical, or quantum) to the outputs $y$. In this model, the relative contribution of the quantum part was $\alpha_q = 0.45(3)$, which meant that the VQC had an active participation in the inference. 

\subsubsection{Performance on a QPU} \label{sec:analysis:qpu}
The hybrid model was trained on the QMware hybrid quantum simulator~\cite{benchmarking}, and tested on the 25-qubit ion-based quantum computer, \textit{IonQ Aria 1}~\cite{ionq_aria} on a short path with three decision points.  The hybrid model was loaded onto AWS Braket and executed on the QPU through PennyLane's AWS integration~\cite{AWS_PL}. The circuit was transpiled into $861$ lines of \texttt{OpenQASM3} code~\cite{openqasm3}, and each task was run for $1000$ shots. Provided that the device was available, tasks took $9\pm1$ minutes, which includes the queuing and transpilation time, but the extent of the latter effects is unclear. 

\begin{figure}
    \centering
    \includegraphics[width=1.0\linewidth]{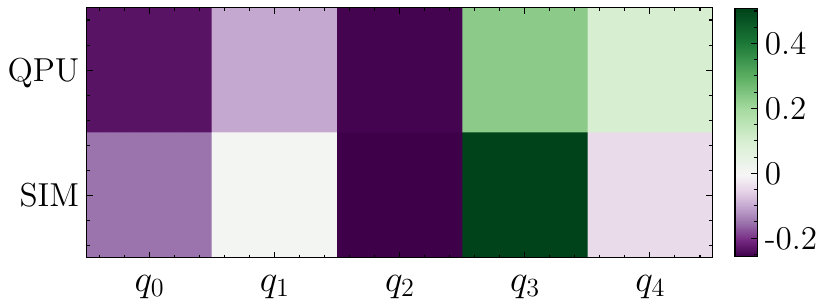}
    \caption{The matrix of the VQC outputs for a decision point generated using the IonQ Aria 1 (QPU) and the QMware cloud simulator (SIM).}
    \label{fig:qpu}
\end{figure}

Fig.~\ref{fig:qpu} compares the VQC outputs for the first decision between the QPU and the infinite-shot simulator. 
Both results show qualitatively the same structure in the activations, and the correlation between the numerically exact simulator and the  actual hardware is rather high. 
However, the concrete values are slightly different, as expected, due to shot noise and gate errors. 
These QPU outputs indicate that the hybrid FiLM model can be executed on today's physical quantum hardware for short paths with few predictions.

\begin{figure*}[t]            
\centering 
     \includegraphics[width=1.0\textwidth]{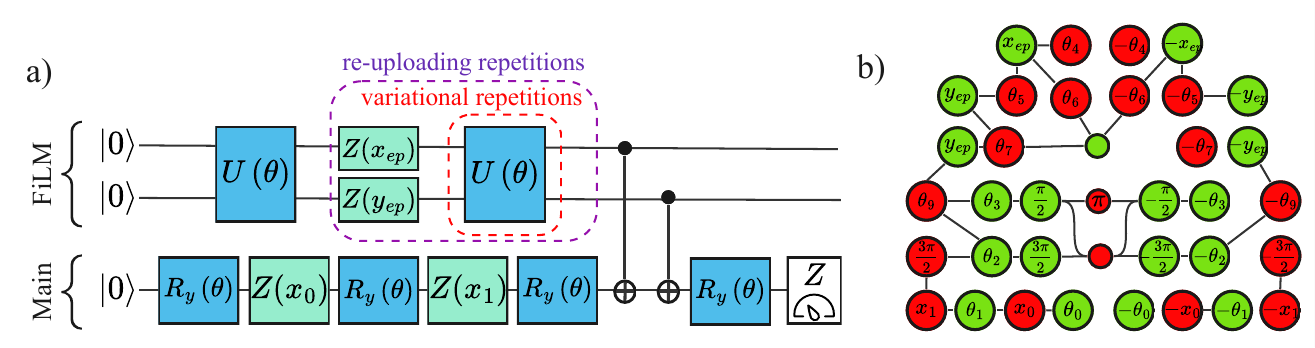}
    \caption{(a) shows the simplified version of the FiLM quantum circuit, with the first two qubits encoding the earthquake coordinates $(x_{\text{ep}},y_{\text{ep}})$ and the third qubit encodes the two main graph features $(x_0,x_1)$. The variational layers in the FiLM section remain the basic entangler layers, but in the main section, they are replaced with Pauli-Y rotations for simplicity.  (b) shows the reduced graph of the example circuit in ZX form for two reuploading repetitions and one variational repetition.}
    \label{fig:ZX-diagram}
\end{figure*}

\subsection{Theoretical Analysis}\label{sec:analysis:theory}

This section theoretically analyses the FiLM quantum model of Fig.~\ref{fig:approaches_SL}. We present a simplified version of the quantum FiLM network with two qubits acting in its FiLM layer and one main circuit qubit (three total qubits) for illustration purposes. Fig.~\ref{fig:ZX-diagram}(a) shows the scaled-down circuit. Analogous to the original circuit, the simplified FiLM quantum network is divided into two parts: the first involves two qubits and operates on earthquake coordinates. The second part utilizes one qubit to handle the main (non-FilM) features which in this case will be two values (instead of the 34 in the exact problem). These two parts are interconnected through CNOT gates. This simplification enables us to perform a qualitative analysis, as the original architecture can be computationally demanding to investigate. We focus on several distinct methodologies in our research, including the ZX-calculus~\cite{zx-calculus} to explore circuit reducibility, Fourier accessibility~\cite{schuld_fourier} to examine the data embedding strategy, and Fisher information~\cite{amirapaper} to investigate the trainable parametrization of the circuit.  It is noteworthy that the exact results will be different compared with the results of analyzing the larger model as this model uses Pauli-Y trainable rotations instead of the BEL layer and fewer qubits in total. Still, the overall results will be indicative of patterns that can be generalized to the model in Fig.~\ref{fig:approaches_SL}.  

\subsubsection{ZX-Calculus Reduction}\label{sec:analysis:zx}

ZX-calculus is a graphical language that replaces circuit diagrams with ZX-diagrams by replacing quantum tensors with so-called ``spiders'', nodes on a graph with edges that connect them ~\cite{zx-calculus, vandewetering2020zxcalculus, wang2023completeness}. These spiders come in two flavors, a light or green-colored spider that represents tensors in the $Z$ basis ($\ket{0}$, $\ket{1}$) and a dark or red-colored spider that represents tensors in the $X$ basis ($\ket{+}$, $\ket{-}$). Once created, ZX diagrams can be simplified and reduced with the language's graphical rewrite rules based on the underlying quantum operations. For example, repetitions of Pauli rotations sum together to form one Pauli rotation with an angle equal to the sum of its parts. This is translated into ZX-calculus as a specific instance of the more general rule of ``fusing'' spiders, where nodes of the same color combine and sum their angles. More generally, quantum operations often possess subtle symmetries that make it difficult to implement effective circuits, and for exponentially large systems, matrix multiplication quickly becomes unwieldy. Essentially, ZX calculus replaces tedious matrix multiplication of quantum gates with easy-to-apply graphical rules. Thus, analysis of ZX-diagrams is helpful for identifying redundancies in a quantum model.

\begin{figure*}[t]            
\centering    
     \includegraphics[width=1.0\linewidth]{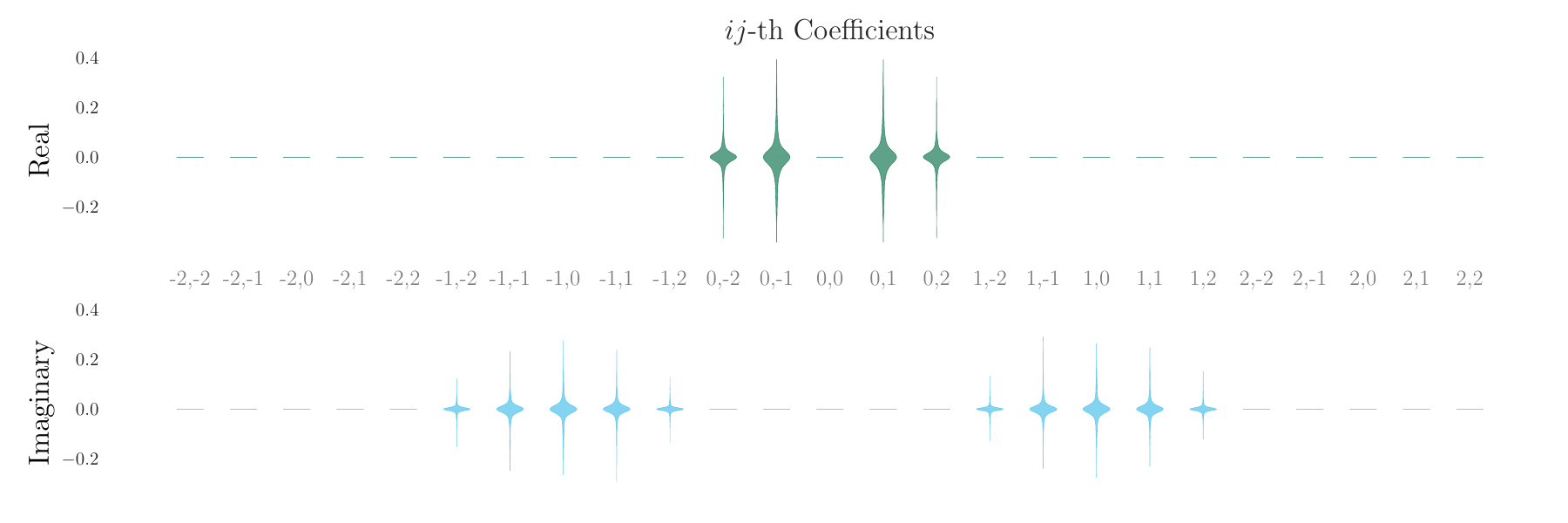}
    \caption{
    (a) shows a violin chart of 1000 samples of the values of the Fourier coefficients for various $\theta$s. The $ij$th indices along the center line
    represent the Fourier coefficient $c_{ij}$. The width of the violins represents the number of samples at that magnitude. The $c_{00}$ term is a global offset and was removed for visual clarity but ranges from $-1$ to $1$. }
    \label{fig:fourier_violin}
\end{figure*}

To analyze the reduced quantum FiLM circuit of  Fig.~\ref{fig:ZX-diagram}(a), we first represented it as a red-green ZX-diagram. Then ZX-calculus's rewriting rules are applied to simplify the circuit and remove redundancies. Finally, the resulting new circuit is extracted. Fig.~\ref{fig:ZX-diagram}(b) shows the simplified circuit and measurement in ZX form. Note that some trainable parameters could be removed in this way. In this example, the trainable parameter labeled `$\theta_8$' -- corresponding to RX rotation applied to the first qubit in the final $U(\theta)$ layer -- is not present as it self-annihilates with its adjoint and thus has no effect on the model output. In other words, the parameterized gate with this weight should be removed.  In the same way, the larger circuit used in this work might include some redundancies that are automatically removed due to ZX analysis. 
    
\subsubsection{Fourier Expressivity} \label{sec:analysis:fourier}

Ref.~\cite{schuld_fourier} showed that the output of a parameterized quantum circuit is equivalent to a truncated Fourier series. It proved that reuploading the features of the dataset $d$ times creates models that approximate the dataset by a Fourier series with degree $d$ -- see Ref.~\cite{exponential,shin} for non-linear scaling with the number of data reuploading layers. For a feature vector of length $N$, the Fourier series as a function of the feature vector $\mathbf{x}$ and trainable parameters $\mathbf{\theta}$ is:
\begin{align*}
f_\mathbf{\theta}(\mathbf{x}) = \sum_{\omega_1 \in \Omega_1} \ldots \sum_{\omega_N \in \Omega_N} c_{\omega_1 \ldots \boldsymbol{\omega}_N} (\mathbf{\theta}) e^{-i \mathbf{\omega} \cdot \mathbf{x}},
\end{align*}
where $\omega_i \in \{ -d_i, \ldots, 0 , \dots, d_i\}$. In other words, the number of terms in the Fourier series is one more than twice the number of times that input was placed in the circuit, $d$. In this analysis, we show the expressivity of the function $f_\mathbf{\theta}(\mathbf{x})$ by sampling over a uniform distribution of random values for each $\theta_i$ from $[0, 2 \pi ]$ and by sampling equidistant $x$ values with a sampling frequency of $d$.

For visual clarity, we display a scaled-down version of the model and only focus on the impact of this sampling on the Fourier terms produced by the FiLM part of the circuit that encodes the earthquake coordinates, namely the coordinates of the earthquake $x$ and $y$. We rewrite $f_\mathbf{\theta}(\mathbf{x}) $ as
\begin{align*}
f_\mathbf{\theta}(x,  y) = \sum^2_{\omega_x =-2} \sum^2_{\omega_y =-2} c_{\omega_x,\omega_y} (\mathbf{\theta}) \: e^{-i \omega_x x  } e^{-i \omega_y y}
\end{align*}

Fig.~\ref{fig:fourier_violin} demonstrates a violin plot of the Fourier coefficients $c_{\omega_x, \omega_y}$ sampled over various $\theta$ realisations. A completely non-expressive model would have terms close to zero for all these coefficients. Instead, the figure shows that the quantum FiLM model is expressive up to 2 coefficient pairs in the real numbers and four coefficient pairs in the imaginary numbers. For the $x$ coefficients, the $d=0$ terms are exclusively real, while the $d=1$ are exclusively imaginary. The final $d=2$ has no expressivity. This is expected since the ZX-calculus in Fig.~\ref{fig:ZX-diagram}(b) shows that one repetition is redundant. Every data reuploading layer is expected to add a new frequency to the degree $d$ for each input, and the expressivity increases as the variational repetitions increase. One could consult the Fisher information matrix to determine the appropriate range for these repetitions.

\subsubsection{Fisher Information} \label{sec:analysis:fisher}

The concept of Fisher information plays a crucial role in comprehending model capacity and trainability \cite{Tan_2021, ly2017tutorial}. Fisher information quantifies the knowledge gained by a statistical model, be it classical or quantum, based on a specific parameterization. In the context of supervised machine learning, we define a family of models with different parameterizations as $\mathcal{F} :={P(\mathbf{x}, \mathbf{y}|\boldsymbol{\theta}): \boldsymbol{\theta} \in \boldsymbol{\Theta}}$, where $(\mathbf{x},\mathbf{y})$ represents the features and targets from our dataset, respectively. Previous studies \cite{amari1998natural, amirapaper} have demonstrated that $\mathcal{F}$ can be viewed as a Riemannian manifold, and the Fisher information matrix can be interpreted as a metric on this manifold, which in some cases (including in this work) can coincide with the loss landscape. The Fisher information matrix is defined as:
\begin{align*}
&F_{ij}(\boldsymbol{\theta}) = \\ 
&\mathbb{E}_{\mathbf{x}, \mathbf{y}}\left[\Big(\partial_{\theta_i} \log P(\mathbf{x}, \mathbf{y}| \boldsymbol{\theta})\Big) 
\Big(\partial_{\theta_j} \log P(\mathbf{x}, \mathbf{y}| \boldsymbol{\theta})\Big)^T\right]
\end{align*}
where the parameters $ \boldsymbol{\theta}$ are the trainable parameters of the HQNN.
\begin{figure*}[!ht]            
\centering  
     \includegraphics[width=1.0\textwidth]{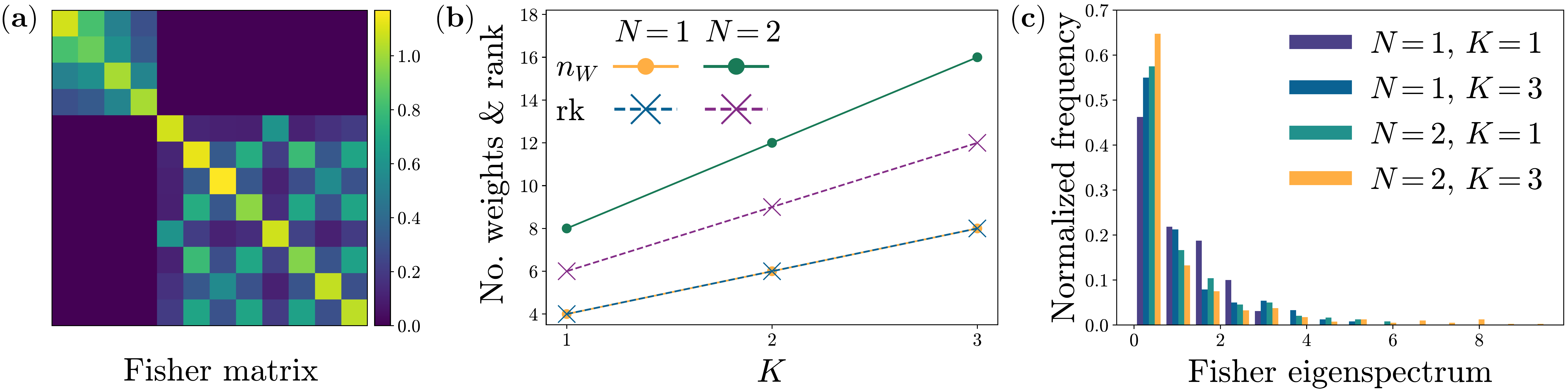}
    \caption{Fisher analysis of the miniaturized quantum FiLM circuit. (a) shows the average Fisher information matrix for a model with $N={1}$ repetition and $K={3}$ reuploadings, considering both circuit parts (FiLM and Main). The average Fisher matrix in this configuration is non-degenerate, and the FiLM and Main weights are independent. (b) demonstrates the relationship between the number of model weights $n_W$(corresponding to the maximal possible rank) and the rank of the average Fisher information matrix across different repetitions ($N=\{1, 2\}$) and reuploadings ($K=\{1, 2, 3\}$). The absence of maximal rank in networks with $N={2}$ indicates the presence of zero gradient parameters. (c) is the Fisher information matrix normalized eigenspectrum frequency for models with $N=\{1, 2\}$ repetitions and $K=\{1, 3\}$ reuploadings. The degeneracy around zero signifies lower trainability.}
    \label{fig: Fisher info}
\end{figure*}

The Fisher information enables us to assess the magnitude of changes in the joint distribution as we traverse the landscape of model parameters. 
As the quantum circuit in this work is a bipartite system with FiLM features  and main features entering separate processing sections, one would reasonably expect a clear separation between the Fisher information matrices of the FiLM and the main sections of the circuit. This hypothesis is confirmed by the Fisher information matrix of the entire circuit shown in Fig.~\ref{fig: Fisher info}(a), where two clearly separated blocks are visible.

The section that processes main features operates with four trainable parameters for a single qubit. The number of trainable parameters for the FiLM earthquake part is larger. It depends on two setup choices: 1) the number of internal variational layers making each variational layer more expressive, and 2) the number of external data reuploading layers increasing the Fourier expressivity of the FiLM layer -- see Sec.~\ref{sec:analysis:fourier}. To explore the variations of the FiLM part, we examine the different combinations of internal layer repetitions, denoted as $N=\{1, 2\}$, and  external reuploading layers, denoted as $K=\{1, 2, 3\}$ showcased in Fig.~\ref{fig:ZX-diagram}(a). Consequently, this configuration's total trainable parameters equals $n_p = 2N(K+1)$. We compute the Fisher information for each setting to assess the architecture configurations. We use $20$ feature samples generated from a Gaussian distribution $\mathbf{x_i} \sim \mathcal{N} (\mu = 0, \sigma^2 = 1)$, with targets encompassing all possible 3-qubit basis states. The Fisher information matrix is calculated using 20 realizations of uniform weights $\theta \in [0, 2\pi)$.

The spectrum of the Fisher information matrix, which encompasses all weights, provides insights into the squared gradients for each parameter \cite{amirapaper}. A network with high trainability will exhibit fewer eigenvalues near zero. Fig.~\ref{fig: Fisher info}(b) illustrates the relationship between the number of trainable parameters and the rank of the Fisher information matrix for different variations of architecture for the circuit earthquake part. We see  that for $N=2$ repetitions, the rank of the Fisher matrix is consistently lower than the number of weights across all reuploading counts. This indicates degeneracy in the Fisher information matrix, implying the presence of zero gradients for certain network parameters. Specifically, a network with $N=2$ repetitions and $K=3$ reuploadings would be expected to have four zero gradient weights. For networks with a single repetition ($N=1$), degeneracy does not occur for all possible reuploadings ($K$). However, the degenerate Fisher matrix is not the primary concern for trainability. Generally, trainability is unaffected if we encounter zero gradients for only a few parameters while the remaining parameters exhibit non-vanishing gradients.
Nonetheless, even networks with non-degenerate Fisher matrices can suffer from low trainability if a significant proportion of eigenvalues have small absolute values. Fig.~\ref{fig: Fisher info}(c) approximates the eigenvalue distribution for each configuration. The frequency of eigenvalues close to zero increases from $46\%$ for $N=1$ repetition and $K=1$ reuploading to nearly $65\%$ for $N=2$ repetitions and $K=3$ reuploadings. Hence, we observe that the likelihood of encountering eigenvalues close to zero tends to rise with the number of trainable parameters.

\subsection{Analysis discussion}
The practical analysis showed that both the quantum, as well as the classical processing, contribute on the same order of magnitude to the output. 
This implies that model primacy is not present in the current scenario. Additionally, we could show a route for running the current architecture on a trapped ion quantum hardware device.      

The ZX calculus reduction shows that stacking the BEL layers can lead to reducible trainable parameters and that efficient training should consider reducing the complete circuit and removing any non-contributing trainable parameters. The Fourier analysis empirically shows the effect of the data reuploading layers on the function-fitting capabilities of the quantum FiLM layer.  It shows that more data reuploading layers leads to an increased number of Fourier terms expressed by the model, and this can be generalized to the full model in Fig.~\ref{fig:approaches_SL}.  Finally, the Fisher matrix shows the clear separation of the trainable parameters of the FiLM and main layers, as well as the increased expressivity and decreased trainability of the model as the BEL sub-layers increase.

\section{Conclusion}\label{sec:conclusion}

This work explored the potential of supervised hybrid quantum machine learning in optimizing emergency evacuation routes during natural disasters. We presented a hybrid supervised learning approach and tested it in a dynamic environment with a dynamic earthquake and increasing traffic congestion at exit points. We compared the approach to a baseline node-wise Dijkstra's algorithm, which finds the shortest path at every new node and time step  given the current situation. Our results showed that hybrid supervised learning could learn to match Dijkstra's algorithm with having access to only a limited portion of the graph, making it a workable option in an uncertain and evolving situation. 
The hybrid model outperforms the purely classical approach and provides an improvement of $7\%$ for the  average accuracy of the model. Additionally, we showed that the quantum part of the hybrid model contributed significantly to the inference.
This study aimed to show the entire stack of hybrid quantum machine learning applied to a practical problem by showcasing: problem formulation,  classical resolution, hybrid formulation and resolution, quantum analysis and efficiency, and QPU integration.  

Overall, the hybrid supervised learning approach has shown promising results in optimizing emergency evacuation plans for cars during earthquakes. Future research needs to focus on testing the approach on different and larger graphs and exploring the potential of other quantum machine learning techniques for solving similar optimization problems in dynamic environments. An exciting avenue for future research could be to examine the potential of reinforcement learning in such a setting. 
This approach could match the promising results of this work and potentially discover even more efficient evacuation paths.

\bibliography{refs.bib}
\end{document}